\documentclass[letterpaper]{article} 
\usepackage{aaai23}  
\usepackage{times}  
\usepackage{helvet}  
\usepackage{courier}  
\usepackage[hyphens]{url}  
\usepackage{graphicx} 
\usepackage{natbib}  
\usepackage{caption} 
\usepackage{algorithm}

\usepackage{multirow}
\usepackage{amsmath}
\usepackage{amssymb}
\usepackage{upgreek}
\usepackage{adjustbox}
\usepackage{makecell}
\usepackage[table,xcdraw]{xcolor}
\usepackage[figuresright]{rotating}
\usepackage{algorithmicx}

\usepackage{amsmath,amssymb,amsfonts}
\usepackage{graphicx}
\usepackage{textcomp}
\usepackage{xcolor}
\usepackage{float}
\usepackage{stfloats}
\usepackage{multirow}
\usepackage{adjustbox}
\usepackage{makecell}
\usepackage{algpseudocode}

\colorlet{red}{black}
\colorlet{blue}{black}
\colorlet{purple}{black}
\colorlet{orange}{black}

\usepackage{newfloat}
\usepackage{listings}
\DeclareCaptionStyle{ruled}{labelfont=normalfont,labelsep=colon,strut=off} 
\floatstyle{ruled}
\newfloat{listing}{tb}{lst}{}
\floatname{listing}{Listing}
\title{M\textsuperscript{3}AE: Multimodal Representation Learning for Brain Tumor Segmentation with Missing Modalities}
\author{
    Hong Liu,\textsuperscript{\rm 1,2,\equalcontrib} 
    Dong Wei,\textsuperscript{\rm 2,\equalcontrib}
    Donghuan Lu,\textsuperscript{\rm 2}
    Jinghan Sun,\textsuperscript{\rm 2,3}
    Liansheng Wang,\textsuperscript{\rm 1,\thanks{Corresponding author.}}
    Yefeng Zheng\textsuperscript{\rm 2}
}
\affiliations{
    \textsuperscript{\rm 1}School of informatics, Xiamen University, Xiamen, China\\
    \textsuperscript{\rm 2}Tencent Jarvis Lab, Tencent Healthcare (Shenzhen) Co., Ltd., Shenzhen, China\\
    \textsuperscript{\rm 3}School of Medicine, Xiamen University, Xiamen, China\\  
    {\{liuhong,jhsun\}@stu.xmu.edu.cn}, 
    {lswang@xmu.edu.cn},
    {\{donwei,caleblu,yefengzheng\}@tencent.com},

}

\usepackage{bibentry}

\begin{document}

\maketitle

\begin{abstract}
Multimodal magnetic resonance imaging (MRI) provides complementary information for sub-region analysis of brain tumors.
Plenty of methods have been proposed for automatic brain tumor segmentation using four common MRI modalities and achieved remarkable performance.
In practice, however, it is common to have one or more modalities missing due to image corruption, artifacts, acquisition protocols, allergy to contrast agents, or simply cost.
In this work, we propose a novel two-stage framework for brain tumor segmentation with missing modalities.
In the first stage, a \underline{m}ulti\underline{m}odal \underline{m}asked \underline{a}uto\underline{e}ncoder (M\textsuperscript{3}AE) is proposed, where both random modalities (i.e., modality dropout) and random patches of the remaining modalities are masked for a reconstruction task, for self-supervised learning of robust multimodal representations against missing modalities.
To this end, we name our framework M\textsuperscript{3}AE.
Meanwhile, we employ model inversion to optimize a representative full-modal image at marginal extra cost, which will be used to substitute for the missing modalities and boost performance during inference.
Then in the second stage, a {\color{red}memory-efficient} self distillation is proposed to distill knowledge between heterogenous missing-modal situations while fine-tuning the model for supervised segmentation.
Our M\textsuperscript{3}AE belongs to the `catch-all' genre where a single model can be applied to all possible subsets of modalities, thus is economic for both training and deployment.
Extensive experiments on BraTS 2018 and 2020 datasets demonstrate its superior performance to existing state-of-the-art methods with missing modalities, 
as well as the efficacy of its components.
Our code is available at: \url{https://github.com/ccarliu/m3ae}.
\end{abstract}

\section{Introduction}
\begin{figure}[t]
    \centering
    \includegraphics[width=.95\columnwidth]{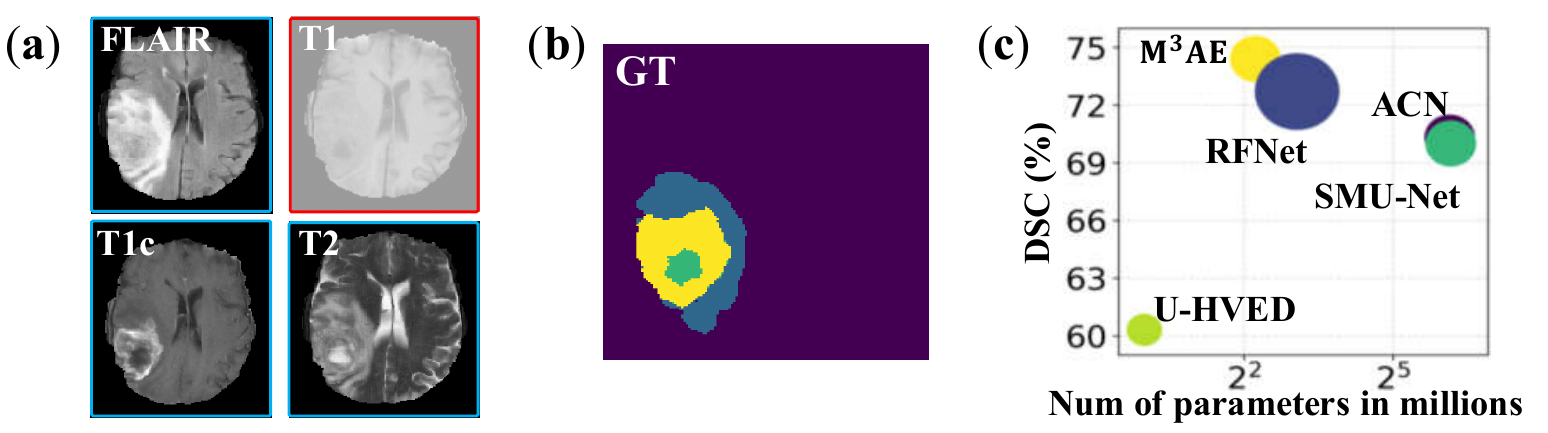} 
    \caption{(a) Example images of the four modalities in BraTS 2018, of which one or more may be missing in practice (e.g., T1 here in red box). (b) Corresponding tumor regions: blue: edema; yellow: enhancing tumor; and green: necrotic and non-enhancing tumor core. (c) Deployment model size (in $\operatorname{log}_2$ scale) and mean Dice similarity scores (DSCs) across all missing-modal situations on BraTS 2018 test split; circle size indicates GFLOPS.
    Compared to four up-to-date methods, our proposed M\textsuperscript{3}AE achieves the best performance with a compact and efficient model.}
    \label{fig:concept}
\end{figure}

Segmentation and associated volume quantification of heterogeneous histological sub-regions are of great value to the diagnosis/prognosis, therapy planning, and follow-up of brain tumors \cite{bakas2018identifying}.
Multi-parametric magnetic resonance imaging (MRI) is the current standard of care for clinical imaging diagnosis of brain tumors \cite{IV201845}.
Specifically, four MRI modalities (in this work, we refer to MRI sequences as modalities) are commonly used to provide complementary information and support sub-region analysis: T1-weighted (T1), contrast enhanced T1-weighted (T1c),  T2-weighted (T2), and T2 fluid attenuation inversion recovery (FLAIR), where the first two highlight tumor core and the last two highlight peritumoral edema (Figs. \ref{fig:concept}(a) and (b)).
In recent years, deep learning methods have greatly advanced the state of the art of brain tumor segmentation with multimodal MRI \cite{chen2020brain,chen2019dual,ding2020multi,myronenko20183d,zhou2020one}. 
However, these methods were optimized for the ideal scenario where the full set of all modalities are present.
While in practice, scenarios of missing one or more modalities commonly occur due to image corruption, artifacts, acquisition protocols, allergy to contrast agents, or simply cost.

To accommodate the practical scenarios of missing modalities, lots of efforts have been made.
A naive approach is to train a `dedicated' model for each possible subset of modalities.
For better performance, the co-training strategy~\cite{blum1998combining} was often incorporated to distill knowledge from full-modal to missing-modal networks \cite{azad2022smu,chen2021learning,hu2020knowledge,wang2021acn}.
Despite their decent performance, the dedicated models were time-costly to train and space-costly to deploy, as $2^N-1$ models were needed for $N$ modalities.
%
Another approach is to synthesize images of missing modalities for full-modal segmentation \cite{2020Assessing,yu2019ea}, where generative adversarial networks \cite[GANs;][]{goodfellow2016nips} were often used.
Notwithstanding, the GANs not only were difficult to train for 3D image generation, but also incurred extra overhead for both training and deployment.
Currently, the predominant approach is to project the available modalities to a common latent space, where a shared feature representation was learned and then projected to the segmentation space \cite{havaei2016hemis,zhou2021feature,zhou2021latent}.
This `catch-all' approach could handle all possible subsets of modalities with a single model, thus was more economic.
However, existing catch-all methods often adopted complex designs with multiple encoders (and sometimes multiple decoders, too) and complicated interactions.

In this work, we propose a novel catch-all framework for brain tumor segmentation using MRI with missing modalities, which features innovative integration of multimodal masked autoencoder, model inversion based modal completion, and memory-efficient self distillation in a single straightforward encoder-decoder architecture.
Above all, witnessing the recent success of masked autoencoders
in learning rich visual representations \cite{he2021masked}, we propose \underline{m}ulti\underline{m}odal \underline{m}asked \underline{a}uto\underline{e}ncoder (M\textsuperscript{3}AE), where a random subset of the modalities and random patches of the remaining ones are masked \emph{simultaneously}.
The intuition is that, to recover the masked content, the model must effectively utilize the inherent inter-modal correlation both globally and locally, plus the intra-modal local semantics.
Accordingly, we name our framework M\textsuperscript{3}AE.
Meanwhile, a representative full-modal image is learned via model inversion \cite{wang2021imagine}, which is served as the substitute for missing modalities during inference and effective in improving performance.
The substitute image is optimized by back propagating the self-supervising M\textsuperscript{3}AE loss, incurring only marginal extra computational cost.
To the best of our knowledge, this work is the first attempt to apply model inversion to modality completion of medical images.
Lastly, we propose a simple yet efficient self distillation \cite{ge2021self,ji2021refine} to promote semantic consistency between different modality combinations.
To this end, we reduce the memory footprint of co-training dual networks, while still able to effectively distill the semantic information between heterogeneous missing-modal situations.
Extensive experiments on two public datasets demonstrate: (1) our framework's robustness to missing modalities and superiority to existing catch-all and dedicated methods (Fig. \ref{fig:concept}(c)),
(2) efficacy of its building components,
and (3) competence of its multimodal representation learning for full modalities.


\section{Related Work}

\subsubsection{Multimodal Brain Tumor Segmentation with Missing Modalities:} 
In this work, we roughly divide existing methods into two categories: dedicated and catch-all.
Several methods proposed to train a dedicated model for each targeted missing situation, where the co-training strategy \cite{blum1998combining} was employed to distill knowledge from full-modal to missing-modal networks.
Both \citet{hu2020knowledge} and \citet{chen2021learning} proposed to distill the knowledge from a multimodal teacher network to monomodal students at the image (i.e., {\color{red}overall semantics}) and pixel (i.e., network output) levels.
Adversarial co-training network \cite[ACN;][]{wang2021acn} enhanced the full- to missing-modal distillation by entropy and knowledge adversarial learning for alignment of the latent representations.
Style matching U-Net \cite[SMU-Net;][]{azad2022smu} decomposed the common latent space of both full- and missing-modal data into content and style representations, and used a  content  and  style-matching  mechanism  to  distill  the informative features from the full-modal network into a missing-modal one.
These methods achieved decent performance especially when more than one modality was missing, while at significant computation and memory costs for both training and deployment ($2^N-1$ models needed for $N$ modalities). 
In contrast, our framework adopts a catch-all design, i.e., a single model applicable to all missing-modal situations, thus is more economic.

A special group of dedicated methods tackled the problem by synthesizing the missing modalities with fidelity \cite{2020Assessing,yu2019ea}, where generative adversarial networks \cite[GANs;][]{goodfellow2016nips} were often used. 
However, GANs are known to be difficult to train for 3D image generation and may incur extra overhead for both training and deployment.
Further, as suggested by \citet{2020Assessing}, the gadolinium contrast agent was indispensable and the resulting contrast images could not be completely reproduced by the generative models.
{\color{purple}
Instead of synthesizing images of missing modalities perfectly for each subject, we opt to optimize a universal full-modal substitute image at marginal cost, which boosts missing-modal segmentation but does not necessarily look realistic.}

The other category of methods attempted to handle all missing-modal situations with a single catch-all model, 
where modality-specific encoders were commonly employed to embed the modalities into a shared latent space, 
followed by feature fusion and further processing to yield segmentation \cite{havaei2016hemis}.
On top of the generic paradigm, hetero-modal variational encoder-decoder \cite[HVED;][]{2019Hetero} incorporated multimodal variational auto-encoders to reconstruct the modalities from the common latent variable, forcing the formulation of a genuinely shared latent representation;
\citet{shen2019brain} proposed adversarial training to adapt feature maps of missing modalities to those of full modalities;
latent correlation representation learning \cite{zhou2021latent} modeled inter-modal correlations to estimate the missing modalities' representation in the latent space;
\citet{zhou2021feature} explicitly generated a feature-enhanced image to provide necessary feature representations of missing modalities; 
and 
region-aware fusion network \cite[RFNet;][]{ding2021rfnet} relied on a region-aware fusion module to conduct feature fusion from available image modalities according to disparate regions adaptively.
All these methods followed complex designs of multiple encoders (and sometimes multiple decoders, too) with complicated interactions.
{\color{purple}Our framework, while also belonging to the catch-all category, is distinct in that it enables a succinct single-encoder-single-decoder architecture (essentially a 3D U-Net) to learn rich multimodal representations and deal with heterogeneous missing-modal situations simultaneously.
}


\subsubsection{\color{purple}Self-Supervised Multimodal Representation Learning for Medical Image Analysis:}
%
While quite a number of works proposed effective self-supervised representation learning for monomodal medical images \cite[][etc.]{taleb20203d,zhang2017self},
researchers have just began to explore unique pretext tasks based on the `multimodality' for multimodal medical images.
Recently, \citet{taleb2021multimodal} introduced a novel cross-modal jigsaw puzzle (CMJP) task to learn a modality-agnostic feature embedding.
Despite its effectiveness, CMJP was proposed for 2D networks and did not consider the practical situations of missing modalities, and how to extend it for 3D networks or missing-modal situations is not straightforward.
The regularizing modality reconstruction task in HVED \cite{2019Hetero} with modality dropout was an effective self-supervising task in preparing for missing modalities.
However, it only focused on global inter-modal correlations but ignored the local structural integrity, which is valuable in learning powerful representation for segmentation.
{\color{purple}In contrast, our M\textsuperscript{3}AE learns rich multimodal representation by modeling both global inter-modal correlations and local intra-modal anatomical integrity, from input with both randomly dropped modalities and randomly masked patches, respectively.
To this end, M\textsuperscript{3}AE is inherently robust to missing modalities while suited for fine-scale semantic learning.}

\begin{figure*}[t]
\centering
\includegraphics[width=.95\textwidth,trim=0 0 0 0,clip]{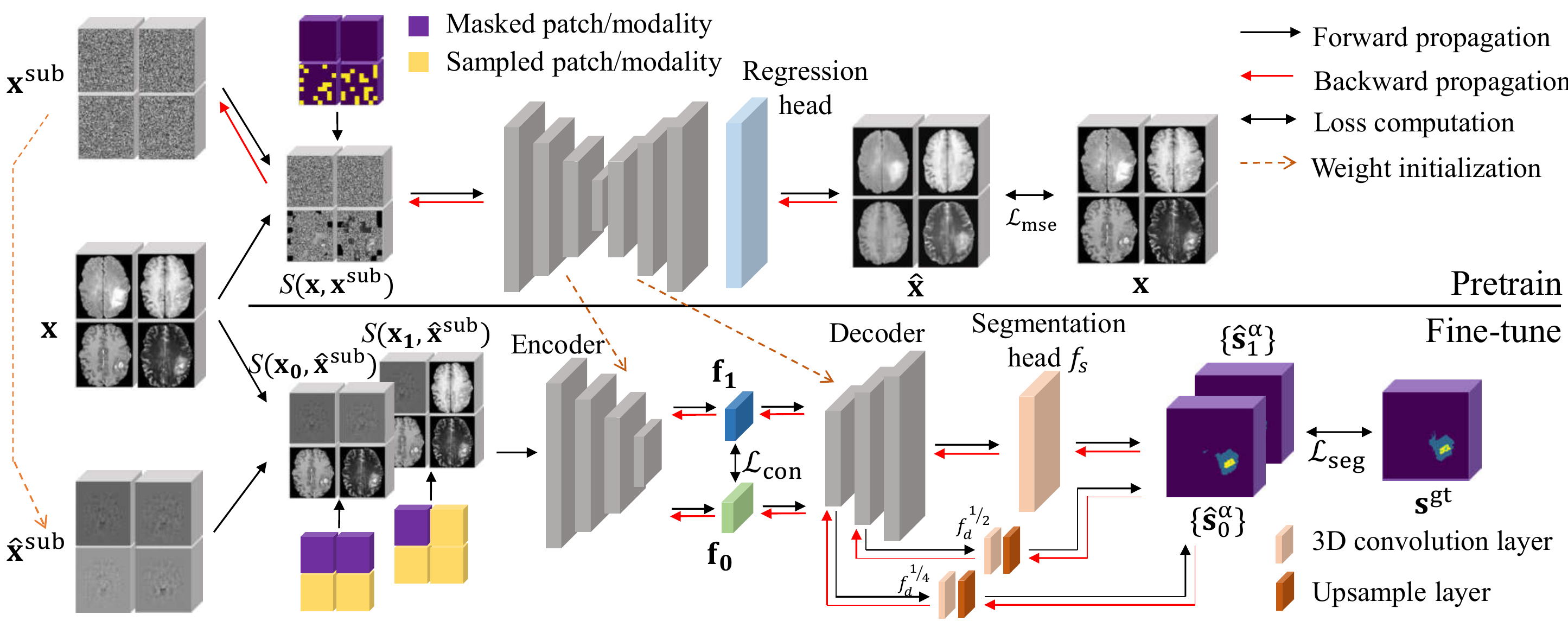}
\caption{Overview of the proposed framework.}\label{fig:overview}
\end{figure*}

\subsubsection{Knowledge Distillation:}
Knowledge distillation \cite[KD;][]{hinton2015distilling} was originally proposed to compress knowledge from one or more teacher networks \textcolor{red}{(often large complex models or model ensemble)} to a student one \textcolor{red}{(often lightweight models)}.
For multimodal segmentation with missing modalities, several works \cite{hu2020knowledge,wang2021acn,chen2021learning,azad2022smu} proposed to transfer the `dark knowledge' of the full-modal network to missing-modal ones via co-training \cite{blum1998combining}.
Although achieving decent performance, the co-training strategy incurred non-negligible memory cost for training due to the dual-network architecture.
In addition, each pair of co-training networks only focused on a fixed correlation between the full modalities and a specific type of missing modalities (e.g., full to T1 alone), failing to exploit the common semantics shared by all different missing-modal situations.
Adopting self distillation \cite{ge2021self,ji2021refine}, our framework distills the shared semantics between \textit{heterogeneous} missing-modal situations (including the special case of full-modal) within a \textit{single} network, and achieving {\color{red} better performance for both missing- and full-modal segmentation} while consuming less resources for training than previous methods.

\section{Method}



The overview of our framework is shown in Fig. \ref{fig:overview}, including a pretraining and a fine-tuning stage.
In the pretraining stage, a novel \underline{m}ulti\underline{m}odal \underline{m}asked \underline{a}uto\underline{e}ncoder (M\textsuperscript{3}AE) is proposed for self-supervised learning of a robust representation against missing modalities.
Meanwhile, a full-modal substitute for missing modalities is learned by model inversion via back propagating the training loss of the M\textsuperscript{3}AE ($\mathcal{L}_\mathrm{mse}$).
Then in the second stage, a {\color{red}memory}-efficient self distillation strategy is proposed to distill the shared semantics between heterogeneous missing-modal situations via a consistency loss ($\mathcal{L}_\mathrm{con}$), while fine-tuning the network for brain tumor segmentation using the supervised loss $\mathcal{L}_\mathrm{seg}$.
The trained segmentation network serves as a `catch-all' model that can be used for any subset as well as the full set of the modalities.
Next, we first describe the newly proposed building components of our framework, including M\textsuperscript{3}AE, model inversion, and self distillation in details, followed by the training and inference procedures integrating them.

\subsubsection{Self-Supervised Multimodal Representation Learning via M\textsuperscript{3}AE:}
Masked autoencoders (MAEs) have been proven successful as scalable self-supervised vision learners \cite{he2021masked}, where the pretext task is to reconstruct the original signal given its partial observation.
Being inspired, we propose a multimodal masked autoencoder (M\textsuperscript{3}AE) for medical images.
Consider a multimodal image $\mathbf{x} \in \mathbb{R}^{N \times D \times H \times W}$, where $W$, $H$, and $D$ are the width, height, and depth of the image, respectively, 
and $N$ is number of modalities.
In practice, any subset of the $N$ modalities can be missing.
Therefore, we sample a random subset of the modalities for masking to mimic the real situation, in addition to randomly masking 3D patches of the remaining modalities as in the original MAE for natural images.
Recovering the modalities masked as a whole requires the network to exploit the global inter-modal correlation, whereas recovering the masked patches requires to exploit both intra-modal structural integrity and local inter-modal correlation.
Thus, our M\textsuperscript{3}AE facilitates self-supervised learning of both anatomical knowledge and inter-modal correlation at the same time.
The mean squared error between the reconstructed and original images ($\hat{\mathbf{x}}$ and $\mathbf{x}$ in Fig. \ref{fig:overview}) is used as the loss function for the M\textsuperscript{3}AE, denoted by $\mathcal{L}_\mathrm{mse}$.

A notable difference between the original MAE for natural images and our M\textsuperscript{3}AE is that, masked patches of the former can only be inferred from surrounding context, whereas those of the latter can be additionally inferred from other modalities and thus expected to be easier.
Therefore, we empirically set an even higher combined masking rate of 87.5\% (compared to 75\% used by \citet{he2021masked}) in our M\textsuperscript{3}AE to make the self-supervising task nontrivial.

\subsubsection{Model Inversion based Modality Completion:}
\begin{figure}
    \centering
    \includegraphics[width=.87\columnwidth,trim=0 5 0 5,clip]{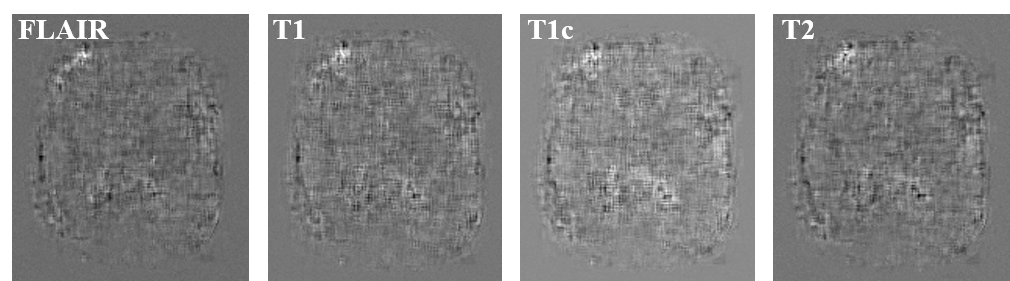}
    \caption{Example full-modal substitute image $\hat{\mathbf{x}}^\mathrm{sub}$ optimized via model inversion on the BraTS 2018 dataset.}\label{fig:x_sub}
\end{figure}

Most existing approaches to modality completion resorted to GANs to synthesize images of the missing modalities \cite{2020Assessing,yu2019ea}, resulting in an extra model and associated training and deployment overheads in addition to the segmentation networks.
Via model inversion, we in this work propose space- and time-efficient synthesis of a full-modal substitute image from the M\textsuperscript{3}AE training process at a marginal cost.
Model inversion has long been used for explainable deep learning, to synthesize images most representative of certain network predictions, e.g., saliency maps for classification \cite{simonyan2014deep}.
Specifically, we optimize an image $\mathbf{x}^\mathrm{sub} \in \mathbb{R}^{N \times D \times H \times W}$ that can lead to smaller reconstruction errors when used to substitute for the masked content (both the whole modalities and intramodal patches) of $\mathbf{x}$:
\begin{equation}\label{eq:inv}
  \hat{\mathbf{x}}^\mathrm{sub} = \operatorname{arg}\min_{\mathbf{x}^\mathrm{sub}} \mathcal{L}_\mathrm{mse}(\mathbf{x}, F(S(\mathbf{x}, \mathbf{x}^\mathrm{sub}))) + \gamma\mathcal{R}(\mathbf{x}^\mathrm{sub}),
\end{equation}
where $S(\mathbf{x}, \mathbf{x}^\mathrm{sub})$ is the operation of replacing the masked content of $\mathbf{x}$ with the location-corresponding content in $\mathbf{x}^\mathrm{sub}$, $F$ is the reconstruction function cascading the backbone network $f$ and a regression head, $\mathcal{R}$ is a regularization term, and $\gamma$ is a weight.
Following \citet{nguyen2016synthesizing}, we use a small amount of $L_2$ regularization for $\mathcal{R}$ with $\gamma=0.005$.
Here, we make a modification to the original MAE \cite{he2021masked} by replacing the masked contents with $\mathbf{x}^\mathrm{sub}$ and processing them in the same way as non-masked ones, instead of discarding them.
The intuition is that, in order to yield better reconstruction, the optimal substitute must capture most representative modality-specific patterns, which are expected to help with the target task of multimodal segmentation, too.
For implementation, $\mathbf{x}^\mathrm{sub}$ is updated by back propagation, along with the update to the network parameters.
In this way, there is no need to introduce any extra module, and the optimization of $\mathbf{x}^\mathrm{sub}$ only incurs marginal cost.
An example of optimized $\hat{\mathbf{x}}^\mathrm{sub}$ is shown in Fig. \ref{fig:x_sub}.
Note that only one $\hat{\mathbf{x}}^\mathrm{sub}$ is learned for a given training dataset, which can be considered as a form of multimodal representation of the training data and is applicable to all subjects.


\subsubsection{Fine-Tune with Heterogeneous Missing-Modal Self Distillation for Tumor Segmentation:}
Implemented via co-training, knowledge distillation from the full-modal to the missing-modal network has proven effective in multimodal segmentation with missing modalities \cite{hu2020knowledge,wang2021acn,chen2021learning,azad2022smu},
although at the great cost of substantial computational overhead due to the pairing network.
%
Inspired by the self distillation strategy \cite{ge2021self,ji2021refine}, we propose a memory-efficient self distillation strategy to distill knowledge between heterogeneous missing-modal situations within a single network.
{\color{purple}Specifically, in each batch, we randomly sample two different missing-modal situations (including the special case of full-modal) of one subject via modality dropout as the network input, and encourage consistent semantic features between them with a consistency loss $\mathcal{L}_\mathrm{con}$:
\begin{equation}\label{eq:distil}
    \mathcal{L}_\mathrm{con}(\mathbf{x}_0,\mathbf{x}_1, \hat{\mathbf{x}}^\mathrm{sub}) =  \mathcal{L}_\mathrm{mse}{(\mathbf{f}_0,\mathbf{f}_1)},
\end{equation}
where $\mathbf{x}_0\text{ and }\mathbf{x}_1$ are the two random missing-modal instantiations of $\mathbf{x}$; $\mathbf{f}_0\text{ and }\mathbf{f}_1\in\mathbb{R}^{C \times D' \times H' \times W'}$ are the corresponding feature maps extracted from $S(\mathbf{x}_0, \hat{\mathbf{x}}^\mathrm{sub})$ and $S(\mathbf{x}_1, \hat{\mathbf{x}}^\mathrm{sub})$, respectively;
and $C$, $D'$, $H'$, and $W'$ are the channel number, depth, height, and width of the feature maps, respectively.
The mutual knowledge transfer via Eqn. (\ref{eq:distil}) is two-way beneficial:
the knowledge transfer from the more to less modalities encourages recovery of the lost information of the missing modalities, 
and that in the reverse direction (especially from monomodal to multimodal) enhances modality-specific features.
In addition, as $\mathbf{x}_0\text{ and }\mathbf{x}_1$ are obtained by random modality dropout in each epoch, our self distillation transfers knowledge between heterogeneous missing-modal situations, instead of between fixed ones as in paired co-training.}
{\color{red}Following \citet{hu2020knowledge} and \citet{wang2021acn}, we distill in the latent space at the bottleneck of the network (see Fig. \ref{fig:overview} bottom).}



\subsubsection{Training and Inference Procedures:}
A two-stage (pretraining and fine-tuning) training scheme is employed. 
In the first stage, the M\textsuperscript{3}AE is trained with both random modality and random patch replacement, with the substitute image $\mathbf{x}^\mathrm{sub}$ optimized at the same time. 
The optimization problem can be formulated as:
\begin{equation}\label{eq:pretrain}
\min_{F,\mathbf{x}^\mathrm{sub}}\mathcal{L}_\mathrm{mse}(\mathbf{x}, F(S(\mathbf{x},\mathbf{x}^\mathrm{sub}))) + \gamma\mathcal{R}(\mathbf{x}^\mathrm{sub}).
\end{equation}
This stage serves as self-supervised pretraining in which the inherent inter-modal correlation and anatomical integrity are learned along with the full-modal substitute image.
Then in the second stage, two missing-modal volumes $\mathbf{x}_0$ and $\mathbf{x}_1$ are instantiated for each case by randomly dropping zero to three modalities.
Also, the regression head used in the first stage is replaced with a randomly initialized segmentation head $f_s$.
{\color{purple}The optimization problem now becomes:
\begin{equation}\label{eq:fine-tune}\small
    \min_{f,f_s,\{f_d\}} \lambda\mathcal{L}_\mathrm{con}(\mathbf{x}_0,\mathbf{x}_1,\hat{\mathbf{x}}^\mathrm{sub}) + 
    {\sum}_{i=0}^1\mathcal{L}_\mathrm{seg}(\mathbf{s}^\mathrm{gt}, \mathbf{x}_i ,\hat{\mathbf{x}}^\mathrm{sub}),
\end{equation}
where $\lambda$ is a weight, $\mathcal{L}_\mathrm{seg}$ is the segmentation loss with deep supervision \cite{lee2015deeply}:
\begin{equation}\small
    \mathcal{L}_\mathrm{seg}(\mathbf{s}^\mathrm{gt}, \mathbf{x}_i,\hat{\mathbf{x}}^\mathrm{sub}) = {\sum}_{\alpha\in\{1, \frac{1}{2}, \frac{1}{4}\}}\mathcal{L}(\mathbf{s}^\mathrm{gt}, \hat{\mathbf{s}}_i^\alpha),\ i\in\{0,1\},
\end{equation}
where $\mathcal{L}$ is the Dice loss \cite{milletari2016v} plus cross entropy loss commonly used for medical image segmentation, $\mathbf{s}^\mathrm{gt}$ is the segmentation ground truth, and $\hat{\mathbf{s}}_i^\alpha$ is the network prediction for $S(\mathbf{x}_i, \hat{\mathbf{x}}^\mathrm{sub})$ at a specific scale $\alpha$ with respect to the input size, which is upsampled (if needed) to match the size of $\mathbf{s}^\mathrm{gt}$.
Specifically, a 1$\times$1$\times$1 convolution (denoted by $f_d^\alpha$) followed by trilinear upsampling is used to yield the intermediate prediction for $\alpha=\frac{1}{2}$ and $\frac{1}{4}$.}
Therefore, the second stage tunes the network for the target task of multimodal segmentation with missing modalities, with the help of the self distilling consistency loss.
As to inference, we simply substitute $\hat{\mathbf{x}}^\mathrm{sub}$ for the missing modalities (if any), feed the resulting image to the trained model, and obtain the segmentation by $f_s$.
Pseudo code of the above-described procedures {\color{red}(assuming each minibatch consisting of a single \textit{subject})} is shown in Algorithm S1.

\section{Experiments and Results}

\subsubsection{Datasets and Evaluation Metrics:}
We evaluate the proposed framework with two widely used multimodal Brain Tumor Segmentation (BraTS) datasets: BraTS 2018 and 2020  \cite{bakas2018identifying}.
The BraTS datasets comprise multi-contrast MRI exams with four sequences: T1, T1c, T2, and FLAIR.
The scans were preprocessed by the organisers, including skull-stripping, re-sampling to a unified resolution (1 mm\textsuperscript{3}), and co-registration to the same template.
Following the challenge, four intra-tumor structures (edema, enhancing tumor, necrotic and non-enhancing tumor core) are grouped into three tumor regions for evaluation: (1) whole tumor, including all tumor tissues, (2) tumor core, composed of the enhancing tumor, necrotic and non-enhancing tumor core, and (3) enhancing tumor.
The BraTS 2018 and 2020 datasets include 285 and 369 training cases with ground truth publicly available, respectively, for which we follow the splits of 199:29:57 (training:validation:testing) and 219:50:100 cases in \cite{ding2021rfnet}, respectively.
%
%
%
{\color{purple}The model is trained on the training splits and tuned (including hyperparameters and other settings) according to the performance on the validation splits, whereas the testing splits are only used for final model evaluation.}
Following the BraTS challenge, 
we use the Dice similarity coefficient (DSC) and the 95\textsuperscript{th} percentile of the Hausdorff distance (HD95) for performance quantification. 

\begin{table*}[htb]
\setlength{\tabcolsep}{.5mm}
\begin{adjustbox}{width=\textwidth}
\begin{tabular}{cccccccccccccccccccccc}
\hline
\multicolumn{4}{c}{Modality}                  &  & \multicolumn{5}{c}{Whole tumor}                      &  & \multicolumn{5}{c}{Tumor core}                       &  & \multicolumn{5}{c}{Enhancing tumor}                  \\ \cline{1-4} \cline{6-10} \cline{12-16} \cline{18-22}
\small{FLAIR}     & T1        & T1c       & T2        &  & U-HVED & ACN &   SMU-Net & RFNet     & M\textsuperscript{3}AE &  & U-HVED & ACN       &    SMU-Net & RFNet  & M\textsuperscript{3}AE &  & U-HVED  & ACN &     SMU-Net & RFNet   & M\textsuperscript{3}AE \\ \hline
\rowcolor[HTML]{EFEFEF}
$\circ$   & $\circ$   & $\circ$   & $\bullet$ &  & 77.5\scriptsize{$\pm$13.8}$^*$ & 84.8\scriptsize{$\pm$11.6} & 84.3\scriptsize{$\pm$10.0}      & \textbf{85.1}\scriptsize{$\pm$9.1} & 84.8\scriptsize{$\pm$10.0} &                    
& 45.0\scriptsize{$\pm$25.9}$^*$ & 70.4\scriptsize{$\pm$22.9}          & \textbf{70.8}\scriptsize{$\pm$21.6}  &  66.9\scriptsize{$\pm$25.6}    & 69.4\scriptsize{$\pm$22.8} &  
&18.7\scriptsize{$\pm$15.6}$^*$& 42.6\scriptsize{$\pm$25.7} & 43.4\scriptsize{$\pm$25.5}   &43.0\scriptsize{$\pm$27.8}&  \textbf{47.6}\scriptsize{$\pm$25.8}     \\
$\circ$   & $\circ$   & $\bullet$ & $\circ$   &  & 57.9\scriptsize{$\pm$24.1}$^*$ & 75.7\scriptsize{$\pm$16.7} & \textbf{77.0}\scriptsize{$\pm$15.6}$^*$      & 73.6\scriptsize{$\pm$19.6} & 75.8\scriptsize{$\pm$16.2} &                    
& 62.7\scriptsize{$\pm$28.9}$^*$ & 81.7\scriptsize{$\pm$18.8}          & 82.5\scriptsize{$\pm$17.2}  &  80.3\scriptsize{$\pm$21.3}    & \textbf{82.9}\scriptsize{$\pm$16.7} &  
&55.5\scriptsize{$\pm$33.4}$^*$& 69.6\scriptsize{$\pm$28.6} & 69.8\scriptsize{$\pm$27.9}   &67.7\scriptsize{$\pm$30.3}&  \textbf{73.7}\scriptsize{$\pm$26.0}     \\
\rowcolor[HTML]{EFEFEF}
$\circ$   & $\bullet$ & $\circ$   & $\circ$   &  & 56.9\scriptsize{$\pm$20.1}$^*$ & \textbf{75.1}\scriptsize{$\pm$19.0}$^*$ & 74.9\scriptsize{$\pm$17.8} & 74.8\scriptsize{$\pm$16.7}      & 74.4\scriptsize{$\pm$17.2} &                    
& 40.7\scriptsize{$\pm$26.5}$^*$ & 65.2\scriptsize{$\pm$24.2}          & 63.9\scriptsize{$\pm$24.4}  &  65.2\scriptsize{$\pm$24.1}    & \textbf{66.1}\scriptsize{$\pm$23.0} &  
&8.9\scriptsize{$\pm$9.2}$^*$& 37.1\scriptsize{$\pm$25.0} & \textbf{38.8}\scriptsize{$\pm$24.5}            &32.3\scriptsize{$\pm$23.2}&    37.1\scriptsize{$\pm$25.3}\\
$\bullet$ & $\circ$   & $\circ$   & $\circ$   &  & 77.0\scriptsize{$\pm$16.2}$^*$ & 86.1\scriptsize{$\pm$8.4}$^*$ & 86.4\scriptsize{$\pm$8.3}$^*$      & 85.8\scriptsize{$\pm$8.4}$^*$ & \textbf{88.7}\scriptsize{$\pm$5.6} &                    
& 41.3\scriptsize{$\pm$22.8}$^*$ & \textbf{67.9}\scriptsize{$\pm$21.1}          & 61.6\scriptsize{$\pm$22.2}$^*$  &  62.6\scriptsize{$\pm$22.3}$^*$    & 66.4\scriptsize{$\pm$22.2} &  
&17.8\scriptsize{$\pm$13.5$^*$}& \textbf{38.2}\scriptsize{$\pm$23.4} & 36.1\scriptsize{$\pm$23.8}            &35.5\scriptsize{$\pm$26.3}&    35.6\scriptsize{$\pm$26.4}\\
\rowcolor[HTML]{EFEFEF}
$\circ$   & $\circ$   & $\bullet$ & $\bullet$ &  & 81.7\scriptsize{$\pm$10.1}$^*$ & 84.2\scriptsize{$\pm$12.2}$^*$ & 85.6\scriptsize{$\pm$14.0}      & 85.6\scriptsize{$\pm$14.0} &  \textbf{86.3}\scriptsize{$\pm$8.7} &                    
& 74.4\scriptsize{$\pm$19.4}$^*$ & 78.9\scriptsize{$\pm$21.9}$^*$          & 82.3\scriptsize{$\pm$20.7}  &  82.4\scriptsize{$\pm$20.1}    & \textbf{84.2}\scriptsize{$\pm$15.7}&  
&63.3\scriptsize{$\pm$30.9}$^*$& 67.8\scriptsize{$\pm$30.0} & 70.6\scriptsize{$\pm$29.3}  &70.6\scriptsize{$\pm$29.3}&    \textbf{75.3}\scriptsize{$\pm$25.4}    \\
$\circ$   & $\bullet$ & $\bullet$ & $\circ$   &  & 65.8\scriptsize{$\pm$18.7}$^*$ & 75.7\scriptsize{$\pm$18.0} & \textbf{77.7}\scriptsize{$\pm$15.8}$^*$      & 77.5\scriptsize{$\pm$16.6} & 77.2\scriptsize{$\pm$15.6} &                    
& 68.9\scriptsize{$\pm$25.1}$^*$ & 79.8\scriptsize{$\pm$21.2}          & 80.4\scriptsize{$\pm$21.1}$^*$  &  81.3\scriptsize{$\pm$21.4}    & \textbf{83.4}\scriptsize{$\pm$15.9}&  
&59.5\scriptsize{$\pm$30.5}$^*$& 68.6\scriptsize{$\pm$29.5} & 70.5\scriptsize{$\pm$28.7}   &68.5\scriptsize{$\pm$30.5}&    \textbf{74.7}\scriptsize{$\pm$25.5}     \\
\rowcolor[HTML]{EFEFEF}
$\bullet$ & $\bullet$ & $\circ$   & $\circ$   &  & 83.7\scriptsize{$\pm$9.3}$^*$ & 85.4\scriptsize{$\pm$10.7}$^*$ & 84.9\scriptsize{$\pm$10.6}$^*$& \textbf{89.0}\scriptsize{$\pm$5.9} & \textbf{89.0}\scriptsize{$\pm$6.8} &                    
& 51.7\scriptsize{$\pm$20.4}$^*$ & 60.6\scriptsize{$\pm$25.8}$^*$          & 61.0\scriptsize{$\pm$24.4}$^*$  &  \textbf{72.2}\scriptsize{$\pm$19.9}    & 70.8\scriptsize{$\pm$21.8}&  
&16.3\scriptsize{$\pm$13.0}$^*$& 35.0\scriptsize{$\pm$24.4}$^*$ & 36.1\scriptsize{$\pm$23.3}$^*$            &38.5\scriptsize{$\pm$25.3}&    \textbf{41.2}\scriptsize{$\pm$26.4}\\
$\circ$   & $\bullet$ & $\circ$   & $\bullet$ &  & 80.5\scriptsize{$\pm$11.8}$^*$ & 84.0\scriptsize{$\pm$14.8} & 85.1\scriptsize{$\pm$10.2}      & 85.4\scriptsize{$\pm$13.6} & \textbf{86.7}\scriptsize{$\pm$7.0} &                    
& 52.9\scriptsize{$\pm$24.3}$^*$ & 69.7\scriptsize{$\pm$22.3}          & 70.8\scriptsize{$\pm$20.0}  &  71.1\scriptsize{$\pm$23.4}    & \textbf{71.8}\scriptsize{$\pm$19.8}&  
&19.3\scriptsize{$\pm$15.4}$^*$& 42.0\scriptsize{$\pm$25.3} & 43.3\scriptsize{$\pm$25.3}   &42.9\scriptsize{$\pm$27.8}&    \textbf{48.7}\scriptsize{$\pm$27.0}    \\
\rowcolor[HTML]{EFEFEF}
$\bullet$ & $\circ$   & $\circ$   & $\bullet$ &  & 85.2\scriptsize{$\pm$8.8}$^*$ & 85.8\scriptsize{$\pm$10.9}$^*$ & 86.3\scriptsize{$\pm$8.8}$^*$      & 89.3\scriptsize{$\pm$6.1} & \textbf{89.9}\scriptsize{$\pm$5.1} &                    
& 51.4\scriptsize{$\pm$20.0}$^*$ & 66.8\scriptsize{$\pm$22.2}          & 66.9\scriptsize{$\pm$21.0}  &  \textbf{71.8}\scriptsize{$\pm$20.1}    & 70.9\scriptsize{$\pm$22.3}&  
&22.1\scriptsize{$\pm$15.2}$^*$& 40.1\scriptsize{$\pm$25.2} & 41.5\scriptsize{$\pm$24.0}            &\textbf{45.4}\scriptsize{$\pm$27.4}&    \textbf{45.4}\scriptsize{$\pm$27.1}\\
$\bullet$ & $\circ$   & $\bullet$ & $\circ$   &  & 84.0\scriptsize{$\pm$10.2}$^*$ & 85.5\scriptsize{$\pm$14.2}$^*$ & 86.7\scriptsize{$\pm$10.1}$^*$      & 89.4\scriptsize{$\pm$5.7} & \textbf{89.7}\scriptsize{$\pm$5.6} &                    
&71.5\scriptsize{$\pm$19.1}$^*$ & 77.3\scriptsize{$\pm$22.0}$^*$          & 74.8\scriptsize{$\pm$24.0}$^*$  &  81.6\scriptsize{$\pm$19.9}$^*$    & \textbf{84.4}\scriptsize{$\pm$14.7}&  
&61.4\scriptsize{$\pm$30.4}$^*$& 67.2\scriptsize{$\pm$30.3} & 68.2\scriptsize{$\pm$29.2}   &72.5\scriptsize{$\pm$26.9}&    \textbf{75.0}\scriptsize{$\pm$25.2}    \\
\rowcolor[HTML]{EFEFEF}
$\bullet$ & $\bullet$ & $\bullet$ & $\circ$   &  & 85.9\scriptsize{$\pm$8.0}$^*$ & 85.5\scriptsize{$\pm$12.6}$^*$ & 84.4\scriptsize{$\pm$15.6}$^*$      & \textbf{89.9}\scriptsize{$\pm$5.5} & 88.9\scriptsize{$\pm$7.9} &                    
& 74.1\scriptsize{$\pm$17.4}$^*$ & 78.9\scriptsize{$\pm$20.2}$^*$          & 76.8\scriptsize{$\pm$25.0}$^*$  &  82.3\scriptsize{$\pm$19.8}$^*$    & \textbf{84.1}\scriptsize{$\pm$17.6}&  
&61.9\scriptsize{$\pm$30.4}$^*$& 65.8\scriptsize{$\pm$30.6}$^*$ & 66.2\scriptsize{$\pm$31.1}   &71.1\scriptsize{$\pm$28.6}&     \textbf{74.0}\scriptsize{$\pm$26.4}     \\
$\bullet$ & $\bullet$ & $\circ$   & $\bullet$ &  & 86.5\scriptsize{$\pm$7.9}$^*$ & 84.2\scriptsize{$\pm$14.4}$^*$ & 83.8\scriptsize{$\pm$14.8}$^*$      & \textbf{90.0}\scriptsize{$\pm$5.5} & 89.9\scriptsize{$\pm$5.2} &                    
& 56.1\scriptsize{$\pm$20.9}$^*$ & 63.9\scriptsize{$\pm$22.5}$^*$ & 60.4\scriptsize{$\pm$24.8}$^*$  &  \textbf{74.0}\scriptsize{$\pm$20.1}    & 72.7\scriptsize{$\pm$20.0}&  
&22.6\scriptsize{$\pm$15.9}$^*$& 38.3\scriptsize{$\pm$26.0}$^*$ & 35.5\scriptsize{$\pm$23.6}$^*$           &\textbf{46.0}\scriptsize{$\pm$27.4}&     44.8\scriptsize{$\pm$26.2}    \\
\rowcolor[HTML]{EFEFEF}
$\bullet$ & $\circ$   & $\bullet$ & $\bullet$ &  & 87.6\scriptsize{$\pm$7.2}$^*$ & 85.6\scriptsize{$\pm$13.8}$^*$ & 84.4\scriptsize{$\pm$15.5}$^*$      & \textbf{90.4}\scriptsize{$\pm$5.6} & 90.2\scriptsize{$\pm$5.5} &                    
& 75.1\scriptsize{$\pm$17.6}$^*$ & 79.6\scriptsize{$\pm$19.4}$^*$          & 75.4\scriptsize{$\pm$23.2}$^*$  &  82.6\scriptsize{$\pm$19.2}    & \textbf{84.6}\scriptsize{$\pm$15.7}&  
&62.9\scriptsize{$\pm$30.6}$^*$& 66.1\scriptsize{$\pm$30.4}$^*$ & 67.2\scriptsize{$\pm$30.6}$^*$   &73.1\scriptsize{$\pm$26.8}&     \textbf{73.8}\scriptsize{$\pm$26.9}    \\
$\circ$   & $\bullet$ & $\bullet$ & $\bullet$ &  & 82.5\scriptsize{$\pm$9.5}$^*$ & 84.9\scriptsize{$\pm$10.4}$^*$ & 83.2\scriptsize{$\pm$17.6}$^*$      & \textbf{86.1}\scriptsize{$\pm$13.6} & 85.7\scriptsize{$\pm$13.5} &                    
& 75.8\scriptsize{$\pm$17.8}$^*$ & 81.3\scriptsize{$\pm$17.1}$^*$          & 78.8\scriptsize{$\pm$22.9}$^*$  &  82.9\scriptsize{$\pm$20.1}    & \textbf{84.4}\scriptsize{$\pm$17.3}&  
&63.6\scriptsize{$\pm$30.8}$^*$& 67.5\scriptsize{$\pm$30.2} & 70.4\scriptsize{$\pm$28.0}   &70.9\scriptsize{$\pm$29.3}&     \textbf{75.4}\scriptsize{$\pm$25.5}     \\
\rowcolor[HTML]{EFEFEF}
$\bullet$ & $\bullet$ & $\bullet$ & $\bullet$ &  & 88.0\scriptsize{$\pm$6.8}$^*$ & 86.2\scriptsize{$\pm$8.9}$^*$ & 85.4\scriptsize{$\pm$9.4}$^*$      & \textbf{90.6}\scriptsize{$\pm$5.4}$^*$ & 90.1\scriptsize{$\pm$5.6} &                    
& 76.2\scriptsize{$\pm$16.6}$^*$ & 79.3\scriptsize{$\pm$17.5}$^*$  & 78.3\scriptsize{$\pm$18.7}$^*$  &  82.9\scriptsize{$\pm$19.0}    & \textbf{84.5}\scriptsize{$\pm$16.7}&  
&63.0\scriptsize{$\pm$30.7}$^*$& 67.4\scriptsize{$\pm$26.9} & 68.1\scriptsize{$\pm$27.8}   &71.4\scriptsize{$\pm$28.4}&     \textbf{75.5}\scriptsize{$\pm$25.3}     \\ \hline
\multicolumn{4}{c}{Mean}                      &  & 78.7\scriptsize{$\pm$10.3}$^*$ & 83.2\scriptsize{$\pm$4.1}$^*$ & 83.3\scriptsize{$\pm$3.7}$^*$      & 85.5\scriptsize{$\pm$5.7} & \textbf{85.8}\scriptsize{$\pm$5.5} &                    
& 61.2\scriptsize{$\pm$13.4}$^*$ & 73.4\scriptsize{$\pm$7.3}$^*$          & 72.3\scriptsize{$\pm$7.9}$^*$  &  76.0\scriptsize{$\pm$7.2}$^*$    & \textbf{77.4}\scriptsize{$\pm$7.6} &  
&41.1\scriptsize{$\pm$22.7}$^*$& 54.2\scriptsize{$\pm$14.8}$^*$ & 55.0\scriptsize{$\pm$15.5}$^*$   &56.6\scriptsize{$\pm$16.0}$^*$&     \textbf{59.9}\scriptsize{$\pm$16.7} \\
\hline
\end{tabular}
\end{adjustbox}
\caption{Performance comparison (DSC \% in mean{\scriptsize$\pm$std}.) with SOTA methods, including U-HVED \cite{2019Hetero}, ACN \cite{wang2021acn}, SMU-Net \cite{azad2022smu}, and RFNet \cite{ding2021rfnet}, on the testing split of BraTS 2018.
Present and missing modalities are denoted by $\bullet$ and $\circ$, respectively. 
{\color{blue}$*$: $p<0.05$ by Wilcoxon signed rank test for pairwise comparison with our method.}
}\label{tab:brats18}
\end{table*}

\begin{table}[t]
    \centering
    \small
    \setlength{\tabcolsep}{.6mm}
    \begin{adjustbox}{width=\columnwidth}
    \begin{tabular}{c|cccc|c}
    \hline
    Method          & U-HVED & ACN  & SMU-Net & RFNet & M\textsuperscript{3}AE \\ \hline
    Whole tumor     & 80.7\scriptsize{$\pm$9.9}$^*$   & 85.4\scriptsize{$\pm$3.4}$^*$ & 85.3\scriptsize{$\pm$3.2}$^*$    & 86.7\scriptsize{$\pm$5.0}  & \textbf{86.9}\scriptsize{$\pm$4.2} \\
    \rowcolor[HTML]{EFEFEF} 
    Tumor core      & 66.5\scriptsize{$\pm$12.8}$^*$   & 77.9{\scriptsize{$\pm$7.0}} & 77.7\scriptsize{$\pm$6.5}    & 78.2\scriptsize{$\pm$7.7}$^*$  & \textbf{79.1}\scriptsize{$\pm$7.2}          \\
    Enhancing tumor & 46.7\scriptsize{$\pm$22.8}$^*$   & 59.9\scriptsize{$\pm$14.0}$^*$ & 59.7\scriptsize{$\pm$14.3}$^*$    & 59.7\scriptsize{$\pm$15.8}$^*$  & \textbf{61.7}\scriptsize{$\pm$16.3}          \\ \hline
    \end{tabular}
    \end{adjustbox}
    \caption{Performance comparison with SOTA methods (see Table \ref{tab:brats18} for references) on the testing split of BraTS 2020.
    The mean performance (DSC \% in mean{\scriptsize$\pm$std}.) of all modal combinations is shown here (due to space limit the detailed performance of each modal combination is given in supplementary Table S3).
    {\color{blue}$*$: $p<0.05$ by Wilcoxon signed rank test for pairwise comparison with our method.}}
    \label{tab:brats20}
\end{table}

\subsubsection{Implementation:}
The PyTorch framework \cite[1.7.1;][]{paszke2019pytorch} is used for experiments.
{\color{purple}We use the same backbone network as \citet{wang2021acn}, which is essentially a 3D U-Net comprising a single encoder and a single decoder employing residual blocks \cite{he2016deep} and group normalization \cite{wu2018group} (more details provided in the supplement).}
{\color{red}As to the regression (for pretraining) and segmentation heads, 1$\times$1$\times$1 convolutions without and with sigmoid are used, respectively.}
No additional post-processing is conducted.
We use two NVIDIA RTX 2080 Ti GPUs for training, {\color{purple}with a batch size of two volumes, i.e., two volumes of two subjects for pretraining, and two random missing-modal instantiations of a subject for fine-tuning}.
The Adam \cite{kingma2014adam} optimizer {\color{red}
is employed
with an initial learning rate of 0.0003 and a cosine decay scheduler \cite{loshchilov2016sgdr}, for both pretraining (600 epochs) and fine-tuning (300 epochs)}.
To standardize all volumes, we clip the volumes to the [1\textsuperscript{st}, 99\textsuperscript{th}] percentiles of the intensity values followed by min-max scaling, and randomly crop them to a fixed size of 128$\times$128$\times$128 voxels for training.
Side length of the random 3D patches is set to 16 {\color{purple}voxels} following \citet{he2021masked}.
$\mathbf{x}^\mathrm{sub}$ is initialized to Gaussian noise.
%
%
%
Common data augmentation is conducted for training, including: random cropping (from 240$\times$240$\times$155 to 128$\times$128$\times$128 voxels); random intensity shift within {\color{red}[$-$0.1, 0.1]} and scaling within [0.9, 1.1]; and random flipping along the axial, coronal, and sagittal axes with a probability of 0.5.
The weight $\lambda$ and masking ratio are empirically set to 0.1 and 0.875, respectively, based on experimental results on the validation split of BraTS 2018 ({\color{red}see the sensitivity analyses in supplementary Fig. S2)}.
%
%
%

\subsubsection{Comparison with State of the Art (SOTA):}
Table \ref{tab:brats18} and Table \ref{tab:brats20} compare the performance of our framework on the BraTS 2018 and 2020 datasets, respectively, with that of four up-to-date approaches 
to brain tumor segmentation with missing modalities\footnote{\color{orange}We also compare to general-purpose multimodal pretraining methods \cite{geng2022multimodal,poklukar2022geometric} in the supplement.}:
U-Net based HVED \cite[U-HVED;][]{2019Hetero}, 
ACN \cite{wang2021acn}, 
SMU-Net \cite{azad2022smu}, 
and RFNet \cite{ding2021rfnet}, with ACN and SMU-Net being dedicated and the other two being catch-all.
{\color{purple}We reproduce the results of U-HVED, ACN, and SMU-Net on our data splits by running the authors' codes, and those of RFNet by using the model checkpoint provided by the authors\footnote{\color{purple}As we use the data splits in \cite{ding2021rfnet}, it is valid to directly use the authors' checkpoints.
Meanwhile, since we would like to compare the results without post-processing---to be fair to all compared methods, we reproduce the results instead of directly reporting the numbers with post-processing in \cite{ding2021rfnet}.}.}

\begin{figure}[t]
\centering
\includegraphics[width=.95\columnwidth,trim=0 0 0 0,clip]{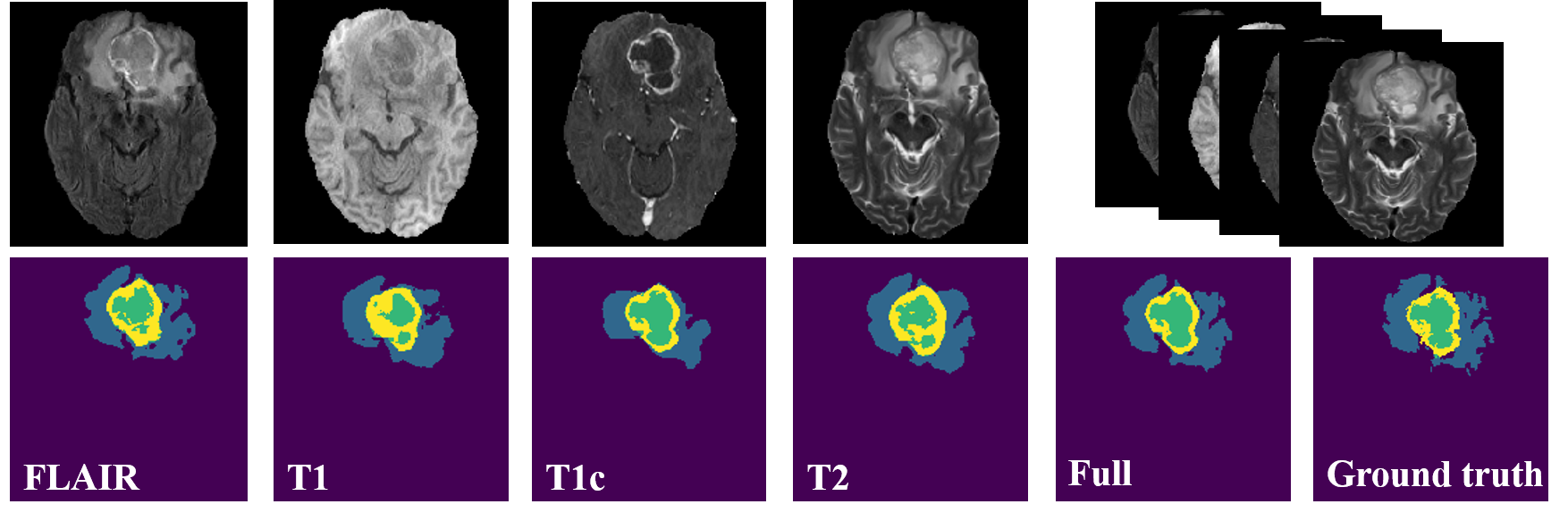}
\caption{Example segmentation results of the proposed M\textsuperscript{3}AE framework on BraTS 2018 using four individual modalities and all of them.
Blue: edema; yellow: enhancing tumor; and green: necrotic and non-enhancing tumor core.}\label{fig:vis}
\end{figure}

As we can see, the proposed M\textsuperscript{3}AE yields the strongest performance for all the three evaluated tumor regions and on both datasets, with the highest mean DSCs averaged over all modal combinations.
It is worth mentioning that as a catch-all method, our M\textsuperscript{3}AE substantially outperforms the two dedicated methods (ACN and SMU-Net), while using only a single trained model.
In contrast, the latter two require 15 models for all the modal combinations.
This makes M\textsuperscript{3}AE more efficient to both train and deploy in practice, in addition to being superior in performance.
We conjecture that two important facts play key roles here:
(1) the co-training strategies employed in the dedicated methods only modelled the one-to-one correlation between the full modalities and each missing-modal situation, whereas our self distillation implicitly models the versatile correlations between all heterogeneous missing-modal situations;
and (2) the random modality dropout and patch masking in M\textsuperscript{3}AE are likely to serve as effective data augmentation that helps model training, which is unavailable in paired co-training where the absent modalities are fixed.
Meanwhile, our M\textsuperscript{3}AE outperforms RFNet, too, which is the previous best performing method and also a catch-all method.
Taking BraTS 2018 for example, compared with RFNet, the mean DSCs of M\textsuperscript{3}AE are slightly higher in whole tumor (85.8\% versus 85.5\%), apparently higher in tumor core (77.4\% versus 76.0\%), and substantially higher in enhancing tumor (59.9\% versus 56.6\%).
Besides, using a vanilla encoder-decoder architecture, our M\textsuperscript{3}AE framework is also more memory-economic and computation-efficient to deploy than RFNet,
which employed multiple encoders with {\color{red}substantially} more parameters and GFLOPS (Fig. \ref{fig:concept}(c)).
{\color{purple}In conclusion, the M\textsuperscript{3}AE framework sets a new SOTA for multimodal brain tumor segmentation with missing modalities, while at the same time using an efficient and economic  architecture for deployment.}
Figure \ref{fig:vis} shows example segmentation results by M\textsuperscript{3}AE.

\begin{table*}[t]
\centering
\setlength{\tabcolsep}{1.2mm}
\small
\begin{adjustbox}{width=\textwidth}
\begin{tabular}{c|cc|ccc|c|cccc|cccc}
\hline
Ablation               & \multicolumn{2}{c|}{Pretrain}                           & \multicolumn{3}{c|}{Completion}      & Self       & \multicolumn{4}{c|}{DSC (\%) $\uparrow$}                      & \multicolumn{4}{c}{HD95 (mm) $\downarrow$}                \\ \cline{2-6}\cline{8-15}
config.                & Mod.Drop                              & M\textsuperscript{3}AE & Mean       & Zero       & Mod.Inv.   & distil.    & Whole         & Core          & Enhancing     & Average       & Whole        & Core         & Enhancing    & Average      \\ \hline
\rowcolor[HTML]{EFEFEF} 
(a)                    & \cellcolor[HTML]{EFEFEF}$\times$ & $\times$             & $\times$   & $\times$   & \checkmark & \checkmark & 85.7\scriptsize{$\pm$5.1}$^*$          & 79.9\scriptsize{$\pm$5.3}$^*$          & 53.1\scriptsize{$\pm$14.5}$^*$          & 72.9\scriptsize{$\pm$6.7}$^*$          & 7.4\scriptsize{$\pm$2.0}          & 9.5\scriptsize{$\pm$2.3}$^*$          & 7.9\scriptsize{$\pm$1.8}$^*$          & 8.2\scriptsize{$\pm$1.6}$^*$          \\
(b)                    & \checkmark                       & $\times$             & $\times$   & $\times$   & \checkmark &  \checkmark  &    \textbf{86.8\scriptsize{$\pm$4.3}}$^*$     &     80.8\scriptsize{$\pm$4.6}$^*$    &      56.8\scriptsize{$\pm$15.9}$^*$      &   74.8\scriptsize{$\pm$6.9}$^*$     &     7.1\scriptsize{$\pm$1.3}    &   9.1\scriptsize{$\pm$2.4}$^*$   &    5.9\scriptsize{$\pm$2.9}    &       7.4\scriptsize{$\pm$1.8}       \\
\rowcolor[HTML]{EFEFEF} 
(c)                    & \cellcolor[HTML]{EFEFEF}$\times$ & \checkmark           & $\times$   & \checkmark & $\times$   & \checkmark & 86.6\scriptsize{$\pm$4.4}          & 80.8\scriptsize{$\pm$4.2}$^*$          & 55.9\scriptsize{$\pm$14.0}$^*$          & 74.4\scriptsize{$\pm$6.1}$^*$          & \textbf{6.7}\scriptsize{$\pm$1.3} & 8.5\scriptsize{$\pm$2.2}          & 6.4\scriptsize{$\pm$2.3}          & 7.2\scriptsize{$\pm$1.5}          \\
(d)                    & $\times$                         & \checkmark           & \checkmark & $\times$   & $\times$   & \checkmark & 48.6\scriptsize{$\pm$28.0}$^*$          & 47.9\scriptsize{$\pm$23.8}$^*$          & 39.3\scriptsize{$\pm$20.6}$^*$          & 45.2\scriptsize{$\pm$22.5}$^*$          & 36.0\scriptsize{$\pm$21.1}$^*$         & 36.9\scriptsize{$\pm$24.5}$^*$         & 12.4\scriptsize{$\pm$10.1}$^*$         & 28.4\scriptsize{$\pm$17.2}$^*$         \\
\rowcolor[HTML]{EFEFEF} 
(e)                    & \cellcolor[HTML]{EFEFEF}$\times$ & \checkmark           & $\times$   & $\times$   & \checkmark & $\times$   & 86.3\scriptsize{$\pm$4.8}          & 80.8\scriptsize{$\pm$4.3}          & 58.2\scriptsize{$\pm$15.4}$^*$          & 75.1\scriptsize{$\pm$6.6}$^*$          & 7.5\scriptsize{$\pm$1.3}$^*$          & 9.2\scriptsize{$\pm$2.0}$^*$          & 6.3\scriptsize{$\pm$2.1}          & 7.6\scriptsize{$\pm$1.5}$^*$          \\ \hline
Full & $\times$                         & \checkmark           & $\times$   & $\times$   & \checkmark & \checkmark & 86.5\scriptsize{$\pm$4.2} & \textbf{81.4}\scriptsize{$\pm$7.0} & \textbf{59.7}\scriptsize{$\pm$16.3} & \textbf{75.9}\scriptsize{$\pm$7.9} & 6.9\scriptsize{$\pm$1.4}          & \textbf{8.4}\scriptsize{$\pm$2.1} & \textbf{5.8}\scriptsize{$\pm$2.6} & \textbf{7.0}\scriptsize{$\pm$1.6} \\ \hline
\end{tabular}
\end{adjustbox}
\caption{Ablation studies on effectiveness of our framework's newly proposed components on the validation split of BraTS 2018, by removing or replacing each component from the `Full' model at a time, including: M\textsuperscript{3}AE self-supervised pretraining, model inversion (Mod.Inv.) based modal completion, and self distillation between heterogeneous missing modalities. 
The mean performance of all modal combinations (format: mean{\scriptsize$\pm$std}) is used.
The `Mod.Drop' pretraining refers to removing the patch masking from our M\textsuperscript{3}AE, i.e., using modality dropout \cite{shen2019brain} alone.
{\color{red}$^*$: $p<0.05$ by Wilcoxon signed rank test for pairwise comparison with the full model.}
}\label{tab:ablate}
\end{table*}


\begin{table}[t]
\centering
\setlength{\tabcolsep}{.5mm}
\begin{adjustbox}{width=\columnwidth}
\begin{tabular}{ccccccccc}
\cline{1-8} 
\multirow{2}{*}{Method} & \multicolumn{3}{c}{DSC (\%) $\uparrow$}       &  & \multicolumn{3}{c}{HD95 (mm) $\downarrow$ }           \phantom{M} &  \\ \cline{2-4} \cline{6-8}
                        & Whole         & Core & Enhancing &  & Whole        & Core         & Enhancing      \\ \cline{1-8}
\rowcolor[HTML]{EFEFEF}
U-HVED$^\dagger$                    & 88.2\scriptsize{$\pm$7.8}$^*$          & 76.5\scriptsize{$\pm$23.6}$^*$ & 70.9\scriptsize{$\pm$27.6}      &  & 5.2\scriptsize{$\pm$4.3}$^*$          & 10.5\scriptsize{$\pm$14.5}$^*$          & 6.7\scriptsize{$\pm$15.1}         \\
ACN$^\dagger$                    & 89.9\scriptsize{$\pm$5.9}$^*$          & 83.3\scriptsize{$\pm$18.1}$^*$ & 77.9\scriptsize{$\pm$24.5}      &  & 6.5\scriptsize{$\pm$6.8}$^*$          & 9.4\scriptsize{$\pm$17.6}$^*$          & 4.5\scriptsize{$\pm$12.6}         \\
\rowcolor[HTML]{EFEFEF}
SMU-Net$^\dagger$                    & 90.1\scriptsize{$\pm$7.3}          & 82.3\scriptsize{$\pm$19.9}$^*$ & 79.2\scriptsize{$\pm$21.5}      &  & 6.8\scriptsize{$\pm$13.1}          & 9.5\scriptsize{$\pm$18.3}          & 6.2\scriptsize{$\pm$15.0}$^*$         \\
RFNet$^\dagger$               & 90.1\scriptsize{$\pm$6.2}$^*$          & 84.6\scriptsize{$\pm$17.1} & 77.2\scriptsize{$\pm$24.0}$^*$      &  & 4.8\scriptsize{$\pm$5.9}$^*$ & \textbf{6.9}\scriptsize{$\pm$13.1} & 5.3\scriptsize{$\pm$13.0}$^*$             \\
\rowcolor[HTML]{EFEFEF}
ModGen$^\dagger$ & 90.2\scriptsize{$\pm$6.0} & 82.5\scriptsize{$\pm$19.9}$^*$ & 77.0\scriptsize{$\pm$24.4}$^*$ & & 5.6\scriptsize{$\pm$8.8} & 8.3\scriptsize{$\pm$14.7} & 4.2\scriptsize{$\pm$8.6} \\
CMJP$^\ddag$  & 89.7 & 84.5 & 79.7     &  & NA & NA & NA \\ 
\rowcolor[HTML]{EFEFEF}
M\textsuperscript{3}AE  & \textbf{90.5}\scriptsize{$\pm$4.8} & \textbf{86.1}\scriptsize{$\pm$13.5} & \textbf{81.0}\scriptsize{$\pm$20.5}     &  &   \textbf{4.6}\scriptsize{$\pm$5.9} & \textbf{6.9}\scriptsize{$\pm$13.5} & \textbf{2.6}\scriptsize{$\pm$3.2}         \\ \cline{1-8}
Challenge$^\ddag$               & 90.4          & 86.0 & 81.5      &  & 4.5          & 8.3          & 3.8          \\ \cline{1-8}
\end{tabular}
\end{adjustbox}
\caption{Full-modal performance comparison on the \emph{online} validation set of BraTS 2018, including U-HVED, ACN, SMU-Net, RFNet (see Table \ref{tab:brats18} for references), Models Genesis \cite[ModGen;][]{zhou2019models}, and cross-modal jigsaw puzzle \cite[CMJP;][]{taleb2021multimodal}.
{\color{red}Single-model} performance of the challenge winner \cite{myronenko20183d} is also included for reference.
Best numbers (excluding the challenge entry) are bolded.
$^\dagger$: reproduced based on the authors' codes;
$^\ddag$: provided by the authors; 
{\color{red}$^*$: $p<0.05$ by Wilcoxon signed rank test for pairwise comparison with our method;}
NA: not available. 
Format: mean{\scriptsize$\pm$std.}, if available.
} \label{tab:full}
\end{table}

\subsubsection{Full-Modal Performance:}
To objectively assess the performance of our framework on full modalities, we also evaluate it on the official validation sets of BraTS 2018 and 2020 
online (https://ipp.cbica.upenn.edu/).
%
The models are trained with the same setting as the offline missing-modal evaluation, except that all the 
cases with public ground truth are used for training without further split (note that the official validation data are kept from training to avoid leakage).
The comparison with other methods on BraTS 2018
is shown in Table \ref{tab:full} (that on BraTS 2020 is given in supplementary Table S1 due to page limit), 
including: U-HVED \cite{2019Hetero}, ACN \cite{wang2021acn}, SMU-Net \cite{azad2022smu}, 
RFNet \cite{ding2021rfnet}, 
Models Genesis \cite[ModGen, a generic self-supervised representation learning method for medical image analysis;][]{zhou2019models},
and CMJP \cite{taleb2021multimodal}.
%
Performance of the challenge winners \cite{isensee2020nnu,myronenko20183d} is included, too, for reference.
%
%
Table \ref{tab:full} shows that compared with other non-challenge approaches,
our M\textsuperscript{3}AE achieves the best performance in both evaluation metrics for all three tumor regions.
It is also mostly comparable and sometimes better than the challenge winner, which involved heavy engineering, e.g., exhaustive parameter tuning.
These results indicate that the multimodal representations learned by our framework are not only robust against missing modalities, but also effective with full modalities.


\subsubsection{Ablation Study:}
To validate the efficacy of our framework's novel building components, we conduct ablative experiments where each component is removed or replaced from the complete model.
The results are shown in Table \ref{tab:ablate}. 
In rows (a) and (b), both removing the M\textsuperscript{3}AE pretraining entirely and {\color{red}removing the patch masking (i.e., keeping the modality dropout alone)} result in {\color{red}apparent drops in average performance}, indicating the indispensable effect of M\textsuperscript{3}AE in learning robust representations of both anatomical and multimodal information against missing modalities.
Compared to row (c), substituting our model inversion optimized full-modal image for missing modalities brings obvious improvements in DSCs upon the zero-filling baseline, whereas substituting the mean image of the training set deteriorates the performance sharply (row (d)).
These results suggest that our optimized substitute image captures useful modal patterns that can complement the missing modalities for brain tumor segmentation, although looks less realistic than the population mean.
Lastly, compared to row (e), the proposed framework is modestly better in both evaluation metrics and for all tumor regions, validating the effectiveness of the two-way hetero-modal knowledge distillation.
In addition, the self distillation strategy saves {\color{red}$\sim$4.7 million parameters} compared to co-training with dual networks.

\section{Conclusion}
This work presented M\textsuperscript{3}AE, a new framework for brain tumor segmentation using MRI with missing modalities.
M\textsuperscript{3}AE featured three novel components: multimodal masked autoencoding for self-supervised learning of robust representations against missing modalities, model inversion based modality completion, and memory-efficient self distillation between heterogeneous missing-modal situations.
As a `catch-all' model, M\textsuperscript{3}AE could accommodate all possible combinations of missing modalities with a single trained model.
Extensive experiments on two public benchmark datasets showed that our framework established a new state of the art for  brain tumor segmentation with missing modalities and that it was competent in multimodal representation learning.
In addition, our ablative experiments validated the efficacy of M\textsuperscript{3}AE's three novel components.
In the future, we plan to apply M\textsuperscript{3}AE to {\color{orange}completely different modalities (e.g., MRI and CT) and other benchmarks beyond BraTS.}

\section{Acknowledgments}
This work was supported in part by the National Key R\&D Program of China under Grant 2020AAA0109500/ 2020AAA0109501, and in part by the National Key Research and Development Program of China (2019YFE0113900).

\bibliography{aaai23}

\begin{thebibliography}{42}
\providecommand{\natexlab}[1]{#1}

\bibitem[{Azad, Khosravi, and Merhof(2022)}]{azad2022smu}
Azad, R.; Khosravi, N.; and Merhof, D. 2022.
\newblock SMU-Net: Style matching U-Net for brain tumor segmentation with
  missing modalities.
\newblock \emph{arXiv preprint arXiv:2204.02961}.

\bibitem[{Bakas et~al.(2018)}]{bakas2018identifying}
Bakas, S.; et~al. 2018.
\newblock {Identifying the best machine learning algorithms for brain tumor
  segmentation, progression assessment, and overall survival prediction in the
  BRATS challenge}.
\newblock \emph{arXiv preprint arXiv:1811.02629}.

\bibitem[{Blum and Mitchell(1998)}]{blum1998combining}
Blum, A.; and Mitchell, T. 1998.
\newblock Combining labeled and unlabeled data with co-training.
\newblock In \emph{Proc. Ann. Conf. on Comput. Learning Theory}, 92--100.

\bibitem[{Chen et~al.(2021)Chen, Dou, Jin, Liu, and Heng}]{chen2021learning}
Chen, C.; Dou, Q.; Jin, Y.; Liu, Q.; and Heng, P.~A. 2021.
\newblock Learning with privileged multimodal knowledge for unimodal
  segmentation.
\newblock \emph{IEEE Trans. Med. Imag.}, 41(3): 621--632.

\bibitem[{Chen et~al.(2020)Chen, Qin, Ding, Tian, and Qin}]{chen2020brain}
Chen, H.; Qin, Z.; Ding, Y.; Tian, L.; and Qin, Z. 2020.
\newblock Brain tumor segmentation with deep convolutional symmetric neural
  network.
\newblock \emph{Neurocomputing}, 392: 305--313.

\bibitem[{Chen, Ding, and Liu(2019)}]{chen2019dual}
Chen, S.; Ding, C.; and Liu, M. 2019.
\newblock Dual-force convolutional neural networks for accurate brain tumor
  segmentation.
\newblock \emph{Pattern Recognit.}, 88: 90--100.

\bibitem[{Ding et~al.(2020)Ding, Gong, Zhang, Li, and Qin}]{ding2020multi}
Ding, Y.; Gong, L.; Zhang, M.; Li, C.; and Qin, Z. 2020.
\newblock A multi-path adaptive fusion network for multimodal brain tumor
  segmentation.
\newblock \emph{Neurocomputing}, 412: 19--30.

\bibitem[{Ding, Yu, and Yang(2021)}]{ding2021rfnet}
Ding, Y.; Yu, X.; and Yang, Y. 2021.
\newblock RFNet: Region-aware fusion network for incomplete multi-modal brain
  tumor segmentation.
\newblock In \emph{Proc. IEEE/CVF Int. Conf. Comput. Vis.}, 3975--3984.

\bibitem[{Dorent et~al.(2019)Dorent, Joutard, Modat, Ourselin, and
  Vercauteren}]{2019Hetero}
Dorent, R.; Joutard, S.; Modat, M.; Ourselin, S.; and Vercauteren, T. 2019.
\newblock Hetero-modal variational encoder-decoder for joint modality
  completion and segmentation.
\newblock In \emph{Proc. Int. Conf. MICCAI}, 74--82. Springer.

\bibitem[{Ge et~al.(2021)}]{ge2021self}
Ge, Y.; et~al. 2021.
\newblock Self-distillation with batch knowledge ensembling improves {ImageNet}
  classification.
\newblock \emph{arXiv preprint arXiv:2104.13298}.

\bibitem[{Geng et~al.(2022)Geng, Liu, Lee, Schuurams, Levine, and
  Abbeel}]{geng2022multimodal}
Geng, X.; Liu, H.; Lee, L.; Schuurams, D.; Levine, S.; and Abbeel, P. 2022.
\newblock Multimodal Masked Autoencoders Learn Transferable Representations.
\newblock \emph{arXiv preprint arXiv:2205.14204}.

\bibitem[{Goodfellow et~al.(2014)}]{goodfellow2016nips}
Goodfellow, I.; et~al. 2014.
\newblock Generative adversarial nets.
\newblock \emph{Adv. Neural Inf. Process. Syst.}, 27.

\bibitem[{Havaei et~al.(2016)Havaei, Guizard, Chapados, and
  Bengio}]{havaei2016hemis}
Havaei, M.; Guizard, N.; Chapados, N.; and Bengio, Y. 2016.
\newblock {HeMIS: Hetero-modal image segmentation}.
\newblock In \emph{Proc. Int. Conf. MICCAI}, 469--477. Springer.

\bibitem[{He et~al.(2022)He, Chen, Xie, Li, Doll{\'a}r, and
  Girshick}]{he2021masked}
He, K.; Chen, X.; Xie, S.; Li, Y.; Doll{\'a}r, P.; and Girshick, R. 2022.
\newblock Masked autoencoders are scalable vision learners.
\newblock In \emph{Proc. IEEE/CVF Conf. Comput. Vis. Pattern Recognit.},
  16000--16009.

\bibitem[{He et~al.(2016)He, Zhang, Ren, and Sun}]{he2016deep}
He, K.; Zhang, X.; Ren, S.; and Sun, J. 2016.
\newblock Deep residual learning for image recognition.
\newblock In \emph{Proc. IEEE/CVF Conf. Comput. Vis. Pattern Recognit.},
  770--778.

\bibitem[{Hinton, Vinyals, and Dean(2015)}]{hinton2015distilling}
Hinton, G.; Vinyals, O.; and Dean, J. 2015.
\newblock Distilling the knowledge in a neural network.
\newblock \emph{arXiv preprint arXiv:1503.02531}.

\bibitem[{Hu et~al.(2020)}]{hu2020knowledge}
Hu, M.; et~al. 2020.
\newblock Knowledge distillation from multi-modal to mono-modal segmentation
  networks.
\newblock In \emph{Proc. Int. Conf. MICCAI}, 772--781. Springer.

\bibitem[{Isensee et~al.(2020)Isensee, J{\"a}ger, Full, Vollmuth, and
  Maier-Hein}]{isensee2020nnu}
Isensee, F.; J{\"a}ger, P.~F.; Full, P.~M.; Vollmuth, P.; and Maier-Hein, K.~H.
  2020.
\newblock nnU-Net for brain tumor segmentation.
\newblock In \emph{International MICCAI Brainlesion Workshop}, 118--132.
  Springer.

\bibitem[{Iv et~al.(2018)Iv, Yoon, Heit, Fischbein, and Wintermark}]{IV201845}
Iv, M.; Yoon, B.~C.; Heit, J.~J.; Fischbein, N.; and Wintermark, M. 2018.
\newblock Current clinical state of advanced magnetic resonance imaging for
  brain tumor diagnosis and follow up.
\newblock \emph{Semin. Roentgenol.}, 53(1): 45--61.

\bibitem[{Ji et~al.(2021)Ji, Shin, Hwang, Park, and Moon}]{ji2021refine}
Ji, M.; Shin, S.; Hwang, S.; Park, G.; and Moon, I.-C. 2021.
\newblock Refine myself by teaching myself: Feature refinement via
  self-knowledge distillation.
\newblock In \emph{Proc. IEEE/CVF Conf. Comput. Vis. Pattern Recognit.},
  10664--10673.

\bibitem[{Kingma and Ba(2014)}]{kingma2014adam}
Kingma, D.~P.; and Ba, J. 2014.
\newblock Adam: A method for stochastic optimization.
\newblock \emph{arXiv preprint arXiv:1412.6980}.

\bibitem[{Lee et~al.(2015)Lee, Xie, Gallagher, Zhang, and Tu}]{lee2015deeply}
Lee, C.-Y.; Xie, S.; Gallagher, P.; Zhang, Z.; and Tu, Z. 2015.
\newblock Deeply-supervised nets.
\newblock In \emph{Artif. Intell. Stat.}, 562--570. PMLR.

\bibitem[{Lee, Moon, and Ye(2020)}]{2020Assessing}
Lee, D.; Moon, W.-J.; and Ye, J.~C. 2020.
\newblock Assessing the importance of magnetic resonance contrasts using
  collaborative generative adversarial networks.
\newblock \emph{Nat. Mach. Intell.}, 2(1): 34--42.

\bibitem[{Loshchilov and Hutter(2016)}]{loshchilov2016sgdr}
Loshchilov, I.; and Hutter, F. 2016.
\newblock {SGDR: Stochastic gradient descent with warm restarts}.
\newblock \emph{arXiv preprint arXiv:1608.03983}.

\bibitem[{Milletari, Navab, and Ahmadi(2016)}]{milletari2016v}
Milletari, F.; Navab, N.; and Ahmadi, S.-A. 2016.
\newblock {V-Net: Fully convolutional neural networks for volumetric medical
  image segmentation}.
\newblock In \emph{Proc. Int. Conf. 3D Vis.}, 565--571. IEEE.

\bibitem[{Myronenko(2018)}]{myronenko20183d}
Myronenko, A. 2018.
\newblock {3D MRI brain tumor segmentation using autoencoder regularization}.
\newblock In \emph{Int. MICCAI Brainlesion Workshop}, 311--320. Springer.

\bibitem[{Nguyen et~al.(2016)Nguyen, Dosovitskiy, Yosinski, Brox, and
  Clune}]{nguyen2016synthesizing}
Nguyen, A.; Dosovitskiy, A.; Yosinski, J.; Brox, T.; and Clune, J. 2016.
\newblock Synthesizing the preferred inputs for neurons in neural networks via
  deep generator networks.
\newblock \emph{Adv. Neural Inf. Process. Syst.}, 29.

\bibitem[{Paszke et~al.(2019)}]{paszke2019pytorch}
Paszke, A.; et~al. 2019.
\newblock {PyTorch: An imperative style, high-performance deep learning
  library}.
\newblock \emph{Adv. Neural Inf. Process. Syst.}, 32.

\bibitem[{Poklukar et~al.(2022)Poklukar, Vasco, Yin, Melo, Paiva, and
  Kragic}]{poklukar2022geometric}
Poklukar, P.; Vasco, M.; Yin, H.; Melo, F.~S.; Paiva, A.; and Kragic, D. 2022.
\newblock Geometric Multimodal Contrastive Representation Learning.
\newblock In \emph{Int. Conf. Mach. Learn.}, 17782--17800. PMLR.

\bibitem[{Shen and Gao(2019)}]{shen2019brain}
Shen, Y.; and Gao, M. 2019.
\newblock Brain tumor segmentation on MRI with missing modalities.
\newblock In \emph{Proc. Int. Conf. IPMI}, 417--428. Springer.

\bibitem[{Simonyan, Vedaldi, and Zisserman(2014)}]{simonyan2014deep}
Simonyan, K.; Vedaldi, A.; and Zisserman, A. 2014.
\newblock Deep inside convolutional networks: Visualising image classification
  models and saliency maps.
\newblock In \emph{Workshop at Int. Conf. Learning Representations}.

\bibitem[{Taleb et~al.(2021)Taleb, Lippert, Klein, and
  Nabi}]{taleb2021multimodal}
Taleb, A.; Lippert, C.; Klein, T.; and Nabi, M. 2021.
\newblock Multimodal self-supervised learning for medical image analysis.
\newblock In \emph{Proc. Int. Conf. IPMI}, 661--673. Springer.

\bibitem[{Taleb et~al.(2020)}]{taleb20203d}
Taleb, A.; et~al. 2020.
\newblock {3D self-supervised methods for medical imaging}.
\newblock \emph{Adv. Neural Inf. Process. Syst.}, 33: 18158--18172.

\bibitem[{Wang et~al.(2021{\natexlab{a}})Wang, Li, Singh, Lu, and
  Vasconcelos}]{wang2021imagine}
Wang, P.; Li, Y.; Singh, K.~K.; Lu, J.; and Vasconcelos, N. 2021{\natexlab{a}}.
\newblock {IMAGINE}: Image synthesis by image-guided model inversion.
\newblock In \emph{Proc. IEEE/CVF Conf. Comput. Vis. Pattern Recognit.},
  3681--3690.

\bibitem[{Wang et~al.(2021{\natexlab{b}})}]{wang2021acn}
Wang, Y.; et~al. 2021{\natexlab{b}}.
\newblock {ACN}: Adversarial co-training network for brain tumor segmentation
  with missing modalities.
\newblock In \emph{Proc. Int. Conf. MICCAI}, 410--420. Springer.

\bibitem[{Wu and He(2018)}]{wu2018group}
Wu, Y.; and He, K. 2018.
\newblock Group normalization.
\newblock In \emph{Proc. Eur. Conf. Comput. Vis.}, 3--19.

\bibitem[{Yu et~al.(2019)Yu, Zhou, Wang, Shi, Fripp, and Bourgeat}]{yu2019ea}
Yu, B.; Zhou, L.; Wang, L.; Shi, Y.; Fripp, J.; and Bourgeat, P. 2019.
\newblock {Ea-GANs: edge-aware generative adversarial networks for
  cross-modality MR image synthesis}.
\newblock \emph{IEEE Trans. Med. Imag.,}, 38(7): 1750--1762.

\bibitem[{Zhang, Wang, and Zheng(2017)}]{zhang2017self}
Zhang, P.; Wang, F.; and Zheng, Y. 2017.
\newblock Self supervised deep representation learning for fine-grained body
  part recognition.
\newblock In \emph{IEEE Int. Symp. Biomed. Imaging}, 578--582.

\bibitem[{Zhou et~al.(2020)Zhou, Ding, Wang, Lu, and Tao}]{zhou2020one}
Zhou, C.; Ding, C.; Wang, X.; Lu, Z.; and Tao, D. 2020.
\newblock One-pass multi-task networks with cross-task guided attention for
  brain tumor segmentation.
\newblock \emph{IEEE Trans. Image Process.}, 29: 4516--4529.

\bibitem[{Zhou et~al.(2021{\natexlab{a}})Zhou, Canu, Vera, and
  Ruan}]{zhou2021feature}
Zhou, T.; Canu, S.; Vera, P.; and Ruan, S. 2021{\natexlab{a}}.
\newblock {Feature-enhanced generation and multi-modality fusion based deep
  neural network for brain tumor segmentation with missing MR modalities}.
\newblock \emph{Neurocomputing}, 466: 102--112.

\bibitem[{Zhou et~al.(2021{\natexlab{b}})Zhou, Canu, Vera, and
  Ruan}]{zhou2021latent}
Zhou, T.; Canu, S.; Vera, P.; and Ruan, S. 2021{\natexlab{b}}.
\newblock {Latent correlation representation learning for brain tumor
  segmentation with missing MRI modalities}.
\newblock \emph{IEEE Trans. Image Process.}, 30: 4263--4274.

\bibitem[{Zhou et~al.(2019)}]{zhou2019models}
Zhou, Z.; et~al. 2019.
\newblock {Models Genesis: Generic autodidactic models for 3D medical image
  analysis}.
\newblock In \emph{Proc. Int. Conf. MICCAI}, 384--393. Springer.

\end{thebibliography}

\clearpage

\end{document}